\documentclass{emulateapj} 
\usepackage{apjfonts}
\usepackage{psfig}
\usepackage{mathrsfs}
\usepackage{amssymb}
\bibpunct[]{(}{)}{;}{a}{}{,} 
\begin{document}
\slugcomment{ApJ Letters, in press}
\newcommand{\LeeZinn}{\mathscr{L}}
\def\vmi{\hbox{\it V--I\/}}
\def\bmv{\hbox{\it B--V\/}}
\def\bmi{\hbox{\it B--I\/}}

\shortauthors{M. Catelan et al.} 
\shorttitle{NGC~6388: Deep {\em HST} Photometry}

\title{Deep {\em HST} Photometry of NGC~6388: Age and Horizontal Branch 
       Luminosity\footnote{Based on observations with the NASA/ESA {\em Hubble Space
       Telescope}, obtained at the Space Telescope Science Institute, which is operated
       by the Association of Universities for Research in Astronomy (AURA), Inc., under 
       NASA contract NAS 5-26555, and on observations retrieved from the ESO ST-ECF
       Archive.}}

\author{M.~Catelan\altaffilmark{1}} 

\author{Peter B. Stetson\altaffilmark{2,}\altaffilmark{3,}\altaffilmark{4}}

\author{Barton J. Pritzl\altaffilmark{5}}  

\author{Horace A. Smith\altaffilmark{6}}  

\author{Karen Kinemuchi\altaffilmark{7}}  

\author{Andrew C. Layden\altaffilmark{8}}  

\author{Allen V. Sweigart\altaffilmark{9}}  

\author{R. M. Rich\altaffilmark{10}}  

\altaffiltext{1}{Pontificia Universidad Cat\'olica de Chile, Departamento de 
       Astronom\'\i a y Astrof\'\i sica, Av. Vicu\~{n}a Mackenna 4860, 
       782-0436 Macul, Santiago, Chile; e-mail: mcatelan@astro.puc.cl}

\altaffiltext{2}{Dominion Astrophysical Observatory, Herzberg Institute of Astrophysics,
       National Research Council, 5071 West Saanich Road, Victoria, BC V9E 2E7, Canada;
       e-mail: Peter.Stetson@nrc.gc.ca}

\altaffiltext{3}{Guest Investigator of the UK Astronomy Data Centre.}

\altaffiltext{4}{Guest User, Canadian Astronomy Data Centre, which is operated
       by the Herzberg Institute of Astrophysics, National Research Council of Canada.}

\altaffiltext{5}{Macalester College, 1600 Grand Avenue, Saint Paul, MN 55105} 

\altaffiltext{6}{Dept.\ of Physics and Astronomy, Michigan State University, 
       East Lansing, MI 48824} 

\altaffiltext{7}{Dept. of Physics and Astronomy, University of Wyoming, Laramie, WY 82071}  

\altaffiltext{8}{Department of Physics and Astronomy, Bowling Green State University, 104
       Overman Hall, Bowling Green, OH 43403} 

\altaffiltext{9}{NASA Goddard Space Flight Center, Exploration of the Universe
       Division, Code 667, Greenbelt, MD 20771} 

\altaffiltext{10}{Division of Astronomy, Department of Physics and Astronomy, UCLA, Los Angeles, 
       CA 90095} 

\begin{abstract}
Using the {\em Hubble Space Telescope}, we have obtained the first
color-magnitude diagram (CMD) to reach the main-sequence turnoff of the Galactic
globular cluster NGC~6388.  From a comparison between the cluster CMD
and 47~Tucanae's, we find that the bulk of the stars in these
two clusters have nearly the same age and chemical composition. On the other
hand, our results indicate that the blue horizontal branch and RR Lyrae
components in NGC~6388 are intrinsically overluminous, which must be due to one
or more, still undetermined, non-canonical second parameter(s) affecting a
relatively minor fraction of the stars in NGC~6388. 
\end{abstract}

\keywords{stars: horizontal-branch --- stars: variables: other --- 
  globular clusters: individual (47~Tucanae, NGC~6388, NGC~6441) --- 
  globular clusters: general}

\section{Introduction}
NGC~6388 and NGC~6441 are two of the most intriguing Galactic globular clusters 
(GC's). The integrated-light study by \citet*{mrea93} revealed a strong 
far-UV flux for these metal-rich \citep*[${\rm [Fe/H]} \simeq -0.60$ and
$-0.53$, respectively;][]{h96} bulge GC's. 
The far-UV flux in resolved GC's is 
dominated by hot horizontal branch (HB) stars \citep*[e.g.,][]{bdea95}, 
especially when rare UV-bright post-asymptotic giant branch stars are 
not present. Accordingly, the most likely explanation for the far-UV flux 
in NGC~6388 and NGC~6441 was immediately recognized to be hot HB stars. However, 
while the UV-upturn phenomenon in elliptical galaxies is often attributed to 
blue HB stars \citep[e.g.,][]{bdea95}, no resolved metal-rich GC 
had been known with a blue HB morphology. 
The tendency for metal-rich GC's to have red HB's while metal-poor 
GC's have predominantly blue HB's reflects the ``first parameter''
of HB morphology. As a consequence, the presence of blue HB stars 
in NGC~6388 and NGC~6441 would represent an example of the so-called 
``second-parameter (2$^{\rm nd}$P) phenomenon.'' 

The presence of blue HB stars extending almost as faint in $V$ as the turnoff (TO) 
point in both NGC~6388 and 
NGC~6441 was confirmed by \citet{mrea97}, who presented {\em Hubble Space 
Telescope} ({\em HST}) photometry for both these GC's from the survey 
by \citet{gpea02}. A  
remarkable feature of the published diagrams is the presence of a {\em strongly 
sloped} HB at colors where other GC's have a much more
nearly ``horizontal'' HB. As emphasized by \citet[][, hereafter SC98]{sc98},
such a sloped HB cannot be simply the result of an older age or of enhanced mass
loss along the red giant branch (RGB): while these 
are able to move a star horizontally along the HB, neither is able to
increase the luminosity of a blue HB star compared to the red HB or RR Lyrae
stars. Likewise, while strong differential reddening
might explain the sloping HB of a red HB cluster, it 
obviously cannot produce RR Lyrae and 
blue HB stars. SC98 conclude therefore that 
{\em non-canonical} 2$^{\rm nd}$P candidates must be at play 
in NGC~6388 and NGC~6441.  

However, these conclusions were challenged by \citet[][, hereafter R02]{grea02}, 
who computed models with non-standard values of the chemical abundance and mixing 
length parameter. Some of their models did reveal sloped HB's, but only as a 
consequence of an anomalously {\em faint red HB} (in $V$), together with a blue 
HB having a $V$-band luminosity consistent with the canonical models 
(see, e.g., their Fig.~2). 

Additional insights are provided by stellar variability and 
spectroscopic studies. \citet{nsea94}, \citet{alea99}, 
\citet{pea00,pea01,pea02,pea03}, and \citet{mcea06} have 
shown that the RR Lyrae variable stars in these GC's, which occupy the normally 
``horizontal'' part of the HB, have much longer periods than field RR Lyrae  
of similar metallicity, thus strongly suggesting that they are intrinsically more 
luminous (SC98). Moreover, 
theoretical calculations by \citet{pea02} have shown that, contrary to 
the suggestions by \citet{crea02}, the RR Lyrae components in both clusters cannot 
be explained in terms of evolution away from a position on the blue HB---and 
neither can the sloping nature of the HB be reproduced in this way. On the other  
hand, the first spectroscopic measurements of the gravities of blue HB stars in both 
NGC~6388 and NGC~6441 \citep*{smea99} revealed surface gravities that are 
higher than predicted by even the canonical models, thus arguing 
against an anomalously bright blue HB~+~RR Lyrae component in these clusters. 
However, a recent reassessment of the spectroscopic 
gravities of blue HB stars in NGC~6388 by \citet{ms06} indicates that 
the actual gravities should, in fact, be lower than the canonical values. It 
appears that the 1999 values must have been in error by a substantial amount, 
probably due to unresolved blends in the crowded inner regions of these   
massive \citep[$M_V \sim -9.5$;][]{h96} GC's. 

In an effort to shed light on this puzzling situation, we have made use of the data 
obtained for NGC~6388 in the course of our snapshot {\em HST} program to study stellar 
variability in the cluster, and also of archival data, to produce its deepest-ever 
color-magnitude diagram (CMD). In \S2 we describe this dataset and 
reduction procedures. In \S3 we compare our CMD with 47~Tucanae's (NGC~104). 
We close in \S4 by discussing the implications of our results 
for our understanding of the origin of the peculiar HB morphology of NGC~6388.

\begin{figure}[t]
  \figurenum{1}
  \begin{center}
  \includegraphics*[width=3.4in]{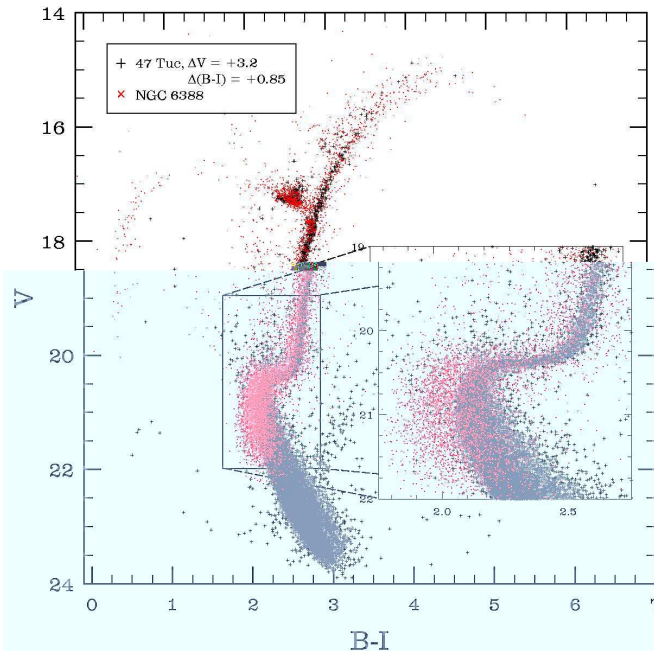}
  \caption{The first deep CMD of NGC~6388 clearly revealing the TO point 
   of the cluster ({\em red $\times$ symbols}) is overplotted on the 47~Tuc CMD 
   ({\em plus signs}) in the $V$, $B\!-\!I$ plane.  
   To produce this plot, the 47~Tuc CMD 
   was shifted by +3.2 in $V$ and by +0.85 in $B\!-\!I$, so as to align their 
   red HB components. 
   }
  \end{center}
      \label{fig:cmd}
\end{figure}

\section{Observational Data and Reduction Procedures} 

The NGC~6388 data used in this paper were obtained under {\em HST} 
program SNAP-9821 (PI B. J. Pritzl), which used the Wide-Field Channel of the {\em
Advanced Camera for Surveys} (ACS) to obtain six $B$ (F435W), $V$ (F555W), $I$ (F814W) 
exposure triptychs on separate dates ranging from 2003
October to 2004 June. In addition, we have retrieved data from the {\em HST}
Archives, as obtained under GO-9835 (PI G. Drukier). These consist of 12 $V$ and
17 $I$ exposures obtained with the High-Resolution Channel of ACS on 2003 October 30.

The results employed here for 47~Tuc were obtained from the ground-based
imagery used by Stetson to define his secondary photometric standards.
They consist of (125, 136, 94) images in ($B$, $V$, $I$) from 12
distinct observing runs; these photometric indices should be on the system of 
\citet{al92} to well under 0.01$\,$mag. Stetson's ground-based data 
also include (84, 137, 99) images in ($B$, $V$, $I$) for NGC~6388.  Although 
these data are not included in the plots in this paper, they were 
used to establish a network of photometric standards in the cluster field which
could be used to establish accurate photometric zero-points for the ACS images.  
Color transformations for the ACS data were based on these local standards and 
similar ACS-ground-based comparisons for 47~Tuc, NGC~2419, NGC~6341 (M92), 
and NGC~6752. 

The data were reduced in the standard manner, using the DAOPHOT-ALLFRAME software 
packages, following commonly understood reduction procedures \citep[e.g.,][]{pbs87,
pbs90,pbs94}.  
A complete description of our dataset, reduction and calibration 
procedures will be described in a future paper (Stetson et al. 2006, in 
preparation). 

Before closing, we note that 
the ACS filter bandpasses are not identical to the standard $B$, $V$, $I$ bandpasses 
of \citet{al92}---but then again, even among the various ground-based observing runs 
where the clusters were observed, the filters and detectors are not identical:  
bandpass mismatch is an unavoidable complication in filter photometry when one 
does not have a privately owned photometer.  Even {\it with\/} a private photometer, 
bandpasses can drift with atmospheric conditions, or as filters, detectors, and mirror 
coatings age.  We have removed the effects of mismatch to second order by including 
linear and quadratic color terms, both in the transformation equations that relate 
our ground-based observations to Landolt's photometric system {\it and\/} in the 
equations that relate the ACS magnitudes to our ground-based system.  Extensive 
experience suggests that the residual effects of bandpass mismatch in these filters 
reach an irreducible minimum scatter of 0.01 -- 0.02 mag on a star-by-star basis, 
and considerably less than this on an ensemble average of many stars.

\begin{figure}[t]
  \figurenum{2}
  \begin{center}
  \includegraphics*[width=3.4in]{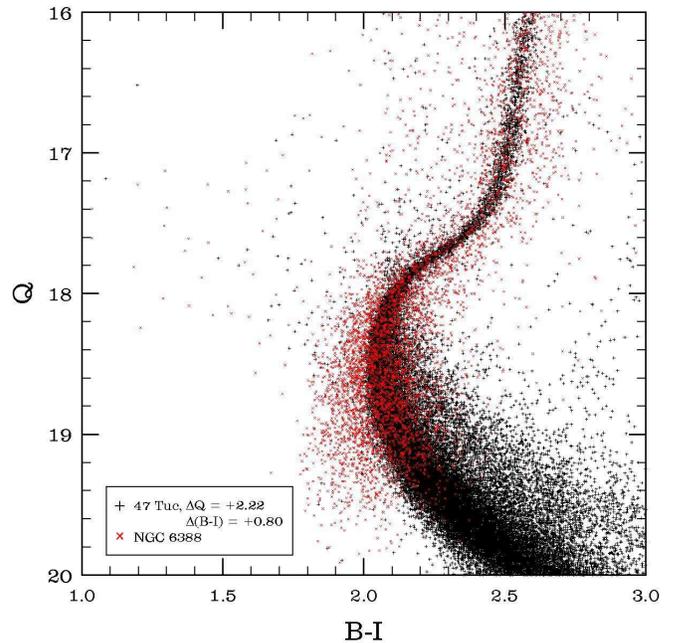}
  \caption{Similar to Figure~\ref{fig:cmd}, but using the reddening-independent
    quantity $Q$ instead of $V$, enforcing a match of the TO colors of the two 
    clusters, and zooming in around the TO/SGB level.  
   }
  \end{center}
      \label{fig:qmd}
\end{figure}

\section{The First Deep CMD of NGC~6388} 

Our deep $V$, $B\!-\!I$ CMD of NGC~6388 is shown in Figure~\ref{fig:cmd} 
({\em red $\times$ symbols}), 
overplotted on the 47~Tuc CMD ({\em plus signs}). 
To obtain this plot, we have applied
shifts of +3.2 in $V$ and +0.85 in \bmi\ to the 47~Tuc 
data, in order to register its red HB to NGC~6388's.  
The magnitude and color of the TO in each cluster were determined by
an iterative robust numerical fit of a parabola to the data in a magnitude range
$\pm 0.5\,$mag of the TO; we obtained ($V$,~\bmi)${}_{\rm TO} = (20.93,~2.04)$ 
for NGC~6388, and (17.70, 1.27) for 47~Tuc. 
Note that we obtained clean, sharp, well-populated sequences
in the CMD by plotting only the stars with the best photometry. 
Our selection criteria will be described in Stetson et al. (2006).

\subsection{The Main Sequence TO and Subgiant Branch (SGB)} 

The TO points and SGB's of 47~Tuc and NGC~6388 
coincide in brightness rather well once their red HB's are 
registered. This suggests 
that any age difference between the two clusters is at the level of 
$\Delta t \lesssim 1$~Gyr, assuming that their red HB's 
have the same absolute luminosity. 

\citet*{pbsea96} and \citet*{dvea98}
computed isochrones for different values of [Fe/H], 
$[\alpha/{\rm Fe}]$, helium abundance $Y$, the mixing length parameter 
$\alpha_{\rm MLT}$ , and rotation rates $\omega$. 
As is already well known, the difference in color between the base of the RGB 
and the main-sequence TO is a good relative age indicator, due to its small 
dependence on [Fe/H], $[\alpha/{\rm Fe}]$, and $\omega$. On the 
other hand, these calculations also show that this color difference, as well 
as the detailed shape of the SGB, do present some dependence on $Y$ and 
$\alpha_{\rm MLT}$.  

To perform a more meaningful differential comparison 
that minimizes the effects of differential reddening 
in NGC~6388, we have replaced $V$ with the reddening-independent quantity 
$Q = V - 1.153 \, (B\!-\!I)$, and then registered the TO points of 
the two clusters in color. Although numerical fitting of parabolas to $Q$ 
as a function of $B\!-\!I$ at the TO's of the two GC's suggests that their 
bluest colors differ by 0.77~mag, we still performed a sanity check by directly 
overplotting the two CMDs with a range of horizontal shifts, and concluded 
that $\Delta(B\!-\!I)\,\sim\,0.80 \pm 0.05$ ($1\sigma$ error bar) 
appeared best to the eye (see Fig.~\ref{fig:qmd}). 
We should note that the detection limit in these data is near $Q\approx19.5$, 
and that past experience \citep[e.g.,][]{sh88} leads 
us to expect that the observed colors may be biased too blue by 
$\sim\,$0.1$\,$mag as this limit is approached. Accordingly, we have been 
careful to restrict our comparisons to the immediate vicinity of the TO 
itself, $Q\,\sim\,18.2$, where any systematic bias is expected to be much 
smaller.

Figure~\ref{fig:qmd} shows that, when the TO points of the clusters are 
registered, the difference in color $\Delta(B\!-\!I)_{\rm TO}^{\rm RGB}$ 
between the base of the RGB and the TO is very similar for both clusters.  
This quantity is only slightly larger for NGC~6388 than for 47~Tuc, at a 
level $\simeq +0.010\pm 0.035$~mag
($1\sigma$ error bar). 
If interpreted in terms of an age difference, this  
translates into $\Delta t \simeq -0.5\pm 1.6$~Gyr (NGC~6388 being 
younger). Assuming that the clusters have similar ages, 
[Fe/H], and $[\alpha/{\rm Fe}]$, which appears most consistent with the 
TO/red HB (``vertical method'') and the RGB data (see also 
the next subsection), one also finds 
$\Delta Y \simeq -0.014\pm 0.024$ (in the sense that NGC~6388 should have 
a {\em lower} $Y$), or $\Delta \alpha_{\rm MLT} \simeq -0.07\pm 0.12$ (again 
in the sense that NGC~6388 should have a lower $\alpha_{\rm MLT}$). 
Such a difference in $Y$ between the two GC's would be the 
opposite of that needed to explain the NGC~6388 HB 
morphology. Stronger constraints on $Y$ and $\alpha_{\rm MLT}$ 
variations between the two clusters are derived from their RGB properties
(see \S3.2).

\subsection{The Red Giant Branch} 
The RGB's of 47~Tuc and NGC~6388 have very similar morphologies, according 
to Figure~\ref{fig:cmd}. The larger scatter in the NGC~6388 CMD than in 
47~Tuc's might be due in part to photometric errors and in part to 
differential reddening, thus making it difficult to determine if the 
stars scattered toward the red of the main RGB in NGC~6388 represent 
a minor metal-rich component. A sizeable metal-poor component 
is clearly not present (see also R02). A small deviation of the bulk 
of the brighter NGC~6388 RGB stars towards redder colors compared with the 47~Tuc 
CMD, if real, would suggest that NGC~6388 might be slightly more metal-rich 
(i.e., by $\lesssim 0.25$~dex) than 47~Tuc 
\citep[${\rm [Fe/H]} = -0.76$~dex;][]{h96}. 
When their red HB's are registered, one also finds that their TO colors become 
somewhat offset (with NGC~6388's being bluer) and that their RGB's actually 
cross just above the HB level (Fig.~\ref{fig:cmd}). Theoretical 
isochrones show that these effects are to be expected if NGC~6388 is more 
metal-rich than 47~Tuc at the level indicated above. 

The ``bump'' in the RGB luminosity function lies at a
$V \approx17.70$ in NGC~6388, and 14.50 in 47~Tuc.
Thus, the TO-to-bump $V$-magnitude differences are 3.26 and 3.20, respectively.  
If real, this difference would suggest that NGC~6388 is 
more metal-{\em poor} (by $\lesssim 0.1$~dex) than 47~Tuc. On the other hand,  
the values of $\Delta V_{\rm bump}^{\rm RHB}$ measured by 
\citet{mzea99} differ by only 0.03~mag between 47~Tuc and NGC~6388; this could be 
produced by a very small metallicity difference, by a $\Delta Y \simeq 0.007$ 
(NGC~6388 being more helium-rich; see Fig.~5 in R02), and/or by a small 
difference in age of $\Delta t \simeq 1$~Gyr (NGC~6388 being younger). 

Note, finally, that near-infrared photometry 
\citep*[e.g.,][]{jfea01,evea04} 
shows that NGC~6388 and NGC~6441 have, if anything, slightly {\em bluer} RGB's 
than 47~Tuc, contrary to what would be expected in the R02 scenario. 
Indeed, \citeauthor{ffea06} find a 
normal value of $\alpha_{\rm MLT} \approx 2.0$ for both NGC~6441 and 47~Tuc 
(NGC~6388 is not in their studied sample). 
The lack of a large difference 
in $\alpha_{\rm MLT}$ and [Fe/H] between the two GC's, while not 
unexpected, already rules out the R02 scenario, in which 
a combination of high metallicity and low $\alpha_{\rm MLT}$ is invoked, 
leading to an underluminous red HB as mentioned in \S1.

\begin{figure}[t]
  \figurenum{3}
  \begin{center}
  \includegraphics*[width=3.4in]{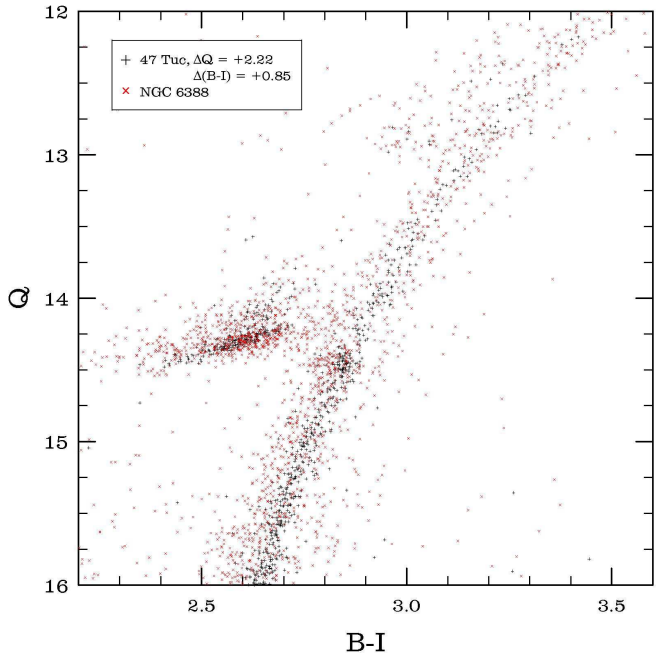}
  \caption{Same as in Figure~\ref{fig:cmd}, but using $Q$ instead of $V$ 
   and focusing around the red HB region. 
   }
  \end{center}
      \label{fig:cmd2}
\end{figure}

\subsection{The Red Horizontal Branch} 
An overluminous red HB, as would be implied by a high primordial 
$Y$ (SC98), is ruled out if, as appears likely, the two clusters have 
similar ages and metallicities. The 
HB luminosity function is also known to be affected by the stars' 
evolutionary parameters; in particular, a higher $Y$ in NGC~6388 
should produce more luminosity evolution away from the zero-age HB 
\citep*[e.g.,][]{bdea89}, which is clearly not present in the 
observed CMD.   Likewise, 
the fact that there is no significant component fainter than the bulk of the 
NGC~6388 red HB stars 
suggests that any metal-rich component in this cluster 
(i.e., with ${\rm [Fe/H]} \gtrsim -0.5$~dex) should be minor. 

Figure~\ref{fig:cmd2} shows a more detailed comparison of the CMDs of 
NGC~6388 and 47~Tuc around the red HB region, again using the 
reddening-independent quantity $Q$. The 47~Tuc $Q$ 
values were shifted by $\Delta Q = +2.22$~mag. 
Adding random scatter at a level of 
$\sigma(B\!-\!I) \simeq 0.03$~mag to the 47~Tuc CMD allows one to reproduce 
the overall appearance of the 
RGB and the red HB quite well, thus implying a $\Delta Y \lesssim 0.03$ 
(in the sense of NGC~6388 having a higher $Y$) based on the luminosity width 
of the HB. The main difference between the two HB's is the tendency for the 
bluer of the NGC~6388 red HB stars to be slightly  
brighter than the average line that defines the 
47~Tuc red HB. This, along with the fact that the TO points and SGB's 
of the two GC's match well in brightness when the two red HB's are 
registered, strongly suggests that both the RR Lyrae and blue HB 
components (and, to a lesser extent, also the bluer red HB stars) 
of NGC~6388 are indeed intrinsically overluminous with respect 
to the 47~Tuc red HB.

\subsection{The Asymptotic Giant Branch ``Clump''} 
Well-populated AGB ``clumps'' are seen in both the 
47~Tuc and NGC~6388 CMD's. \citet{ffea99} have shown that, 
the bluer the HB morphology 
of a GC, the less pronounced the resulting AGB clump. Therefore, 
the majority of the stars in this phase originate from red HB stars. 
It is interesting 
to note that the difference in $V$ magnitude between the AGB clump and the red 
HB is basically indistinguishable between NGC~6388 and 47~Tuc. Unfortunately, 
this does not provide us with strong constraints on the difference in $Y$ 
or metallicity between the two GC's, since the difference 
in magnitude between the AGB clump and the red HB is not very sensitive to 
these evolutionary parameters \citep*[e.g.,][]{gbea95}.

\section{Discussion} 
In the present {\em Letter}, we have shown that, apart from the blue HB  
and RR Lyrae components, the CMD's of NGC~6388 and 47~Tuc are very  
similar, thus strongly suggesting that the bulk of the stars 
in these two clusters are very similar. Differences in age, metallicity, 
$[\alpha/{\rm Fe}]$, $Y$, and $\alpha_{\rm MLT}$ between the two GC's, 
if present, should be small. Our results suggest that 
the red HB component of NGC~6388 is neither underluminous 
(as in the ``canonical tilt'' scenario of R02) nor overluminous (as in the 
high-$Y$ scenario of SC98), except for its bluest stars. 
On the other hand, 
both the RR Lyrae and blue HB stars in NGC~6388 {\em are} significantly 
overluminous compared to field RR Lyrae stars of similar [Fe/H], thus 
explaining NGC~6388's anomalously long RR Lyrae periods. 
The lack of 
a sizeable luminosity difference between the red HB's of NGC~6388 and 47~Tuc 
indicates that {\em the 2$^{\it nd}$P that leads to the production of an 
overluminous RR Lyrae~+~blue HB component in NGC~6388 must be non-canonical 
in nature}---that is, it must be neither age nor RGB mass loss, or else a 
sloped HB would not result (SC98). However, since only a 
relatively small fraction ($\simeq 17\%$; \citeauthor{mc06} \citeyear{mc06})
of the HB stars are in the RR Lyrae strip or on the blue HB, only a minor 
fraction of the cluster stars should be affected by this non-canonical 
2$^{\rm nd}$P, which could be either $Y$ or/and the helium-core mass at the 
He-flash. (A fraction of the red HB stars, especially the bluer 25\% 
or so, could also be affected, though to a much smaller extent.) Additional, 
detailed studies of the blue HB, RR Lyrae, and main sequence components  
will be required before we are in a position to conclusively decide what 
2$^{\rm nd}$P(s) is (are) responsible \citep[see][ for a recent review]{mc06}.

\acknowledgments We thank the referee, G. Piotto, for a very helpful report. 
M.C. acknowledges support by Proyecto FONDECYT Regular 
No. 1030954. B.J.P. would like to thank NASA for the support for the SNAP-9821 
project through a grant from the Space Telescope Science Institute. 
H.A.S. acknowledges support from the CSCE and NSF under AST 02-05813.




\begin{thebibliography}{}

%

\bibitem[Bono et al.(1995)]{gbea95}
  Bono, G., Castellani, V., Degl'Innocenti, S., \& Pulone, L. 1995, \aap, 297, 115



\bibitem[Catelan(2006)]{mc06}
  Catelan, M. 2006, in Resolved Stellar Populations, ASP Conf. Ser., ed. D. 
  Valls-Gabaud \& M. Ch\'avez, in press (astro-ph/0507464)



\bibitem[Corwin et al.(2006)]{mcea06}
  Corwin, T. M., Sumerel, A. N., Pritzl, B. J., Smith, H. A., Catelan, M., 
  Sweigart, A. V., \& Stetson, P. B. 2006, \aj, 132, 1014


\bibitem[Dorman et al.(1995)Dorman, O'Connell, \& Rood]{bdea95}
  Dorman, B., O'Connell, R. W., \& Rood, R. T. 1995, \apj, 442, 105

\bibitem[Dorman et al.(1989)Dorman, VandenBerg, \& Laskarides]{bdea89}
  Dorman, B., VandenBerg, D. A., \& Laskarides, P. G. 1989, \apj, 343, 750

\bibitem[Ferraro et al.(1999)]{ffea99}
  Ferraro, F. R., Messineo, M., Fusi Pecci, F., de Palo, M. A., Straniero, O., 
  Chieffi, A., \& Limongi, M. 1999, \aj, 118, 1738

\bibitem[Ferraro et al.(2006)]{ffea06}
  Ferraro, F. R., Valenti, E., Straniero, O., \& Origlia, L. 2006, \apj, 
  642, 225

\bibitem[Frogel et al.(2001)]{jfea01}
  Frogel, J. A., Stephens, A., Ram\'\i rez, S., \& DePoy, D. L. 2001, \aj, 
  122, 1896


\bibitem[Harris(1996)]{h96}
  Harris, W. E. 1996, \aj, 112, 1487 (Feb.~2003 update)

\bibitem[Holtzman et al.(1995)]{jhea95}
  Holtzmann, J. A., et al. 1995, \pasp, 107, 1065

\bibitem[Landolt(1992)]{al92}
  Landolt, A. U. 1992, \aj, 104, 340

\bibitem[Layden et al.(1999)]{alea99}
  Layden, A. C., Ritter, L. A., Welch, D. L., \& Webb, T. M. A. 1999, \aj, 
  117, 1313

\bibitem[Moehler et al.(1999)Moehler, Sweigart, \& Catelan]{smea99}
  Moehler, S., Sweigart, A. V., \& Catelan, M. 1999, \aap, 351, 519

\bibitem[Moehler \& Sweigart(2006)]{ms06}
  Moehler, S., \& Sweigart, A. V. 2006, \aap, 455, 943


\bibitem[Piotto et al.(2002)]{gpea02}
  Piotto, G., et al. 2002, \aap, 391, 945


\bibitem[Pritzl et al.(2000)]{pea00}
  Pritzl, B., Smith, H. A., Catelan, M., \& Sweigart, A. V. 2000, \apjl, 530, L41

\bibitem[Pritzl et al.(2001)]{pea01}
  Pritzl, B. J., Smith, H. A., Catelan, M., \& Sweigart, A. V. 2001, \aj, 122, 2600

\bibitem[Pritzl et al.(2002)]{pea02}
  Pritzl, B. J., Smith, H. A., Catelan, M., \& Sweigart, A. V. 2002, \aj, 124, 949

\bibitem[Pritzl et al.(2003)]{pea03}
  Pritzl, B. J., Smith, H. A., Stetson, P. B., Catelan, M., Sweigart, A. V., 
  Layden, A. C., \& Rich, R. M. 2003, \aj, 126, 1381

\bibitem[Raimondo et al.(2002)]{grea02}
  Raimondo, G., Castellani, V., Cassisi, S., Brocato, E., \& Piotto, G. 
  2002, \apj, 569, 975 (R02) 

\bibitem[Ree et al.(2002)]{crea02}
  Ree, C. H., Yoon, S.-J., Rey, S.-C., \& Lee, Y.-W. 2002, in Omega Centauri, A Unique 
  Window into Astrophysics, ASP Conf. Ser. 265, ed. F. van Leeuwen, J. D. Hughes, \& 
  G. Piotto (San Francisco: ASP), 101

\bibitem[Rich et al.(1993)Rich, Minniti, \& Liebert]{mrea93}
  Rich, R. M., Minniti, D., \& Liebert, J. 1993, \apj, 406, 489

\bibitem[Rich et al.(1997)]{mrea97}
  Rich, R. M., et al. 1997, \apj, 484, L25 

\bibitem[Silbermann et al.(1994)]{nsea94}
  Silbermann, N. A., Smith, H. A., Bolte, M., \& Hazen, M. L. 1994, \aj, 
  107, 1764

\bibitem[Stetson(1987)]{pbs87}
  Stetson, P. B. 1987, \pasp, 99, 191

\bibitem[Stetson(1990)]{pbs90}
  Stetson, P. B. 1990, \pasp, 102, 932

\bibitem[Stetson(1994)]{pbs94}
  Stetson, P. B. 1994, \pasp, 106, 250



\bibitem[Stetson \& Harris(1988)]{sh88}
  Stetson, P. B., \& Harris, W. E. 1988, \aj, 96, 909 

\bibitem[Stetson et al.(1996)Stetson, VandenBerg, \& Bolte]{pbsea96}
  Stetson, P. B., VandenBerg, D. A., \& Bolte, M. 1996, \pasp, 108, 560


\bibitem[Sweigart \& Catelan(1998)]{sc98}
  Sweigart, A. V., \& Catelan, M. 1998, \apjl, 501, L63 (SC98)

\bibitem[Valenti et al.(2004)Valenti, Ferraro, \& Origlia]{evea04}
  Valenti, E., Ferraro, F. R., \& Origlia, L. 2004, \mnras, 351, 1204

\bibitem[VandenBerg et al.(1998)VandenBerg, Larson, \& De Propris]{dvea98}
  VandenBerg, D. A., Larson, A. M., \& De Propris, R. 1998, \pasp, 110, 98

\bibitem[Zoccali et al.(1999)]{mzea99}
  Zoccali, M., Cassisi, S., Piotto, G., Bono, G., \& Salaris, M. 1999, \apjl, 
  518, L49

\end{thebibliography}
\end{document}